# Comments on the breakdown of the Landauer bound for information erasure in the quantum regime

Elias P. Gyftopoulos
Massachusetts Institute of Technology
77 Massachusetts Avenue, Room 24-111
Cambridge, Massachusetts 02139 USA
and
Michael R. von Spakovsky
Virginia Polytechnic Institute and State University
Blacksburg, Virginia 24061 USA

In Ref. [1], the authors claim that the Landauer principle requires dissipation (release) of at least $T dS$ units of energy as a consequence of erasure of $dS$ units of information, and that they provide counter-examples, based on different models and calculations, which prove that the principle is violated at very low temperatures, and valid at high temperatures.

In our view, neither the Landauer principle nor the models and calculations in Ref. [1] have any relation to either the correct understanding of thermodynamic definitions, postulates, and theorems, or to the affinity that exists between physics and thermodynamics. Physics can be viewed as a large tree with many branches. Thermodynamics is not a branch. It pervades the entire tree.

As shown in a purely thermodynamic discussion of Maxwell's demon [2], none of the hundreds of articles [3], and none of the half a dozen books that have been written about him, including publications by Landauer, have addressed the problem posed by Maxwell. One definitive, nonquantal exorcism of the demon, completely consistent with Maxwell's specifications is given in Ref. [2]. It is based on an unambiguous, noncircular, self-consistent, and nonstatistical exposition of thermodynamics [4] that applies to all systems (both large and small, including one spin systems), and to all states (both thermodynamic equilibrium with inverse varying temperatures either from zero to infinity or from minus infinity to plus infinity, and not thermodynamic equilibrium for which temperature cannot be defined). In addition, in this exposition entropy is shown to be an instantaneous, nonstatistical property of each constituent of any system in any state, in the same sense that inertial mass is an instantaneous, nonstatistical property of each constituent of any system in any state.

Moreover, a nonstatistical, quantum-theoretic unification of mechanics and thermodynamics provides another definitive exorcism of the demon [5], and a quantum analytical expression of entropy [6]. The latter result shows that entropy is a measure of the quantum-mechanical spatial shape of the constituents of a system, in the same sense that $mv^2/2$ and not some other expression is a measure of kinetic energy in classical mechanics, and that spontaneous entropy increase in an isolated system (the only cause of irreversibility) is due to the change of the quantum mechanical spatial shape of the constituents as they try to conform to the applied, and inter-constituent forces of the system [7]. Examples of spatial shapes, shape changes, and entropy increases are discussed in Ref. [8].

Finally, in contrast to entropies of either statistical or informational theories that are interpreted as representing disorder, with the ultimate disorder being achieved if a system of given energy, volume, and amounts of constituents is in a state of thermodynamic equilibrium, both in the nonquantal exposition of thermodynamics and, in the unified quantum theory of mechanics and thermodynamics, each thermodynamic equilibrium state is shown to represent perfect order [9].

It is clear that the preceding results are part of a coherent *complete theory of physical reality*, and contrary to the ad hoc theories and numerical approximations inherent in statistical and informational interpretations of thermodynamics.